\documentclass[12pt]{article}
\usepackage[utf8]{inputenc}
\usepackage[numbers,sort&compress]{natbib}
\usepackage{siunitx}
\usepackage[usenames,dvipsnames,svgnames]{xcolor}
\definecolor{nicered}{rgb}{0.7,0.1,0.1}
\definecolor{nicegreen}{rgb}{0.1,0.5,0.1}
\usepackage[colorlinks=true,citecolor= nicegreen,linkcolor=nicered]{hyperref}
\usepackage[numbers,sort&compress]{natbib}
\usepackage{slashed}
\usepackage{amsmath,amssymb,amsfonts}
\usepackage{graphicx}
\graphicspath{{figures/}}

\usepackage[top=2cm, bottom=3.5cm, left=2cm, right=2cm]{geometry}
\usepackage[affil-it]{authblk}

\newcommand{\orcid}{\includegraphics{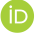}}
\newcommand{\orcidlink}[1]{\href{https://orcid.org/#1}{{\orcid}}}

\title{Type-II two-Higgs-doublet model in noncommutative geometry}

\author[1]{Fredy Jimenez\orcidlink{0000-0002-5175-692X}\footnote{ \href{mailto:fangel.jimenez@udea.edu.co}{fangel.jimenez@udea.edu.co}  }}
\author[1,2]{Diego Restrepo\orcidlink{0000-0001-6455-5564}\footnote{ \href{mailto:restrepo@udea.edu.co}{restrepo@udea.edu.co}  }}
\author[1]{Andrés Rivera\orcidlink{0000-0002-6256-8545}\footnote{ \href{mailto:afelipe.rivera@udea.edu.co}{afelipe.rivera@udea.edu.co}  }}
\affil[1]{Instituto de Física, Universidad de Antioquia,\\
 Calle 70 No 52-21, Medellín, Colombia}
\affil[2]{Instituto de F\'isica Gleb Wataghin, UNICAMP, 13083-859, Campinas, SP, Brazil}

\date{}

\begin{document}

\maketitle
\begin{abstract}
In noncommutative geometry (NCG) the spectral action principle predicts the standard model (SM) particle masses by constraining the scalar and Yukawa couplings at some heavy scale, but gives an inconsistent value for the Higgs mass.
Nevertheless, the scalar sector in the NCG approach to the standard model, is in general composed of two Higgs doublets and its phenomenology remains unexplored. 
In this work, we present a type-II two-Higgs-doublet model  in NCG, with a SM-like Higgs mass compatible with the 125~GeV experimental value and extra scalars within the alignment limit without decoupling with masses from $350$~GeV. 
\end{abstract}

\section{Introduction}
After the discovery of the Higgs scalar particle by the ATLAS and
CMS experiments at the Large Hadron Collider (LHC)~\cite{ATLAS:2012yve,CMS:2012qbp},
the question remains whether the discovered scalar is just one piece
of an extended Higgs sector.
The two-Higgs-doublet model (2HDM)~\cite{Branco:2011iw,Eriksson:2009ws} is
one of the simplest standard model (SM) extensions.
It offers a rich phenomenology, which could explain some of the facts
pointing out the need for physics beyond the SM, like the evidence of
baryon asymmetry~\cite{Trodden:1998qg}, the presence of dark matter
in the universe~\cite{Planck:2018vyg,Camargo:2019ukv} or the smallness of Dirac neutrino masses~\cite{Ma:2021szi}.
It demands the inclusion of two complex doublet scalar fields,
$\Phi_1$ and $\Phi_2$, sharing the same hypercharge, and both acquiring vacuum
expectation values (VEVs), $v_1$ and $v_2$, respectively.
The current measurements of the Higgs boson properties show no
substantial deviations from SM expectations~\cite{ATLAS:2020rej,CMS:2021ugl}, and only one of the neutral Higgs mass eigenstates
`aligns' to the SM-like Higgs boson.
Despite this, searches for additional Higgs bosons in collider experiments provide only weak absolute exclusion limits on the masses
of the scalars~\cite{Haller:2018nnx}, which results in a large freedom
in the choice of the 2HDM parameters.
With this in mind, it would be desirable to have mathematical
structures which could impose more rigid bounds on the parameter space
of these kind of models with extended Higgs sectors.
   
Connes' noncommutative geometry (NCG)~\cite{Connes:1990qp,Connes:1994yd,Connes:1995tu} is a branch of mathematics whose main application to
particle physics is to endow the SM with an internal geometry.
Such an underlying geometrical structure makes difficult to fit SM
extensions into the NCG context
\cite{Krajewski:1996dt,Stephan:2007fa,Squellari:2007zr}. In consequence, it would be desirable  to  use this framework to constraint the parameter space of models with extended Higgs sectors.
In NCG, the topological and the metric data of manifolds are encoded by
spectral triples $\{A, H, D\}$, which are defined by an associative
algebra $A$ represented on a Hilbert space $H$, together with the
action of the Hermitian (Dirac) operator $D$, which
contains the masses and mixing terms for the fermion fields.
This algebra can be considered in general as noncommutative (including
finite-dimensional matrix algebras) so that the Dirac operator allows
to extrapolate differential geometry notions to spaces with a finite
number of points. 
In particular, the SM internal geometry is characterized by the
finite-dimensional (and noncommutative) matrix algebra
$\mathbb{C}\oplus\mathbb{H}\oplus M_3(\mathbb{C})$, where its
automorphism group contains the SM gauge group
SU(3)$\times$SU(2)$\times$U(1).
The faithful representation of this algebra on a $30$-dimensional
(complex) Hilbert space accounts for the SM fermions and antifermions
with their respective quantum numbers.
The action functional describing the dynamics of the
bosonic fields is given by the spectral action principle
\cite{Chamseddine:1996zu}, which is based on the spectrum of the Dirac
operator mixed with the continuous (Riemannian) space-time.
This action constrains the Yukawa and scalar couplings at the grand unification scale, so that by flowing the renormalization group equations
(RGEs) down to the electroweak scale it is possible
to make phenomenological estimations on the masses of the particle fields.
In particular, NCG predicted the existence of a
Higgs boson (provided with a geometric interpretation) as the internal
analog of a gauge field, but with a wrong mass near to $170$ GeV~\cite{Chamseddine:2010ud}.
Such discordance can be solved by enforcing the appearance of a
singlet scalar field in the spectral action~\cite{Chamseddine:2012sw,Khozani:2017ykq}, which can be turned on rigorously by considering a larger algebra~\cite{Devastato:2013oqa,Devastatao:2014xga}, by relaxing the first-order axiom~\cite{Chamseddine:2013rta, Chamseddine:2013kza}, or from a
 $U(1)_{B-L}$ (baryon-lepton) gauge symmetry~\cite{Farnsworth:2014vva}.

In this work, we try to get the proper SM-like Higgs boson mass but preserving the minimal SM algebra and gauge symmetry. Instead of introducing a singlet scalar field, we build a phenomenologically viable 2HDM in the context of NCG. We observe that the Higgs sector (encoded into the inner fluctuations of the Dirac operator) is composed of two complex doublet scalar fields when we use the second-order axiom~\cite{Boyle:2014wba} rather than the massless photon condition~\cite{Chamseddine:2006ep}.
For simplicity, we assume a $CP$-conserving 2HDM scalar potential given by~\cite{Bernon:2015qea}
 \begin{align}
V=&\mu_1^2\Phi_1^{\dagger}\Phi_1+\mu_2^2\Phi_2^{\dagger}\Phi_2- \left(\mu_{12}^2\Phi_1^{\dagger}\Phi_2+\text{h.c.}\right)+\frac{\lambda_1}{2}\left(\Phi_1^{\dagger}\Phi_1\right)^2+\frac{\lambda_2}{2}\left(\Phi_2^{\dagger}\Phi_2\right)^2\nonumber\\
&+\lambda_{3}\Phi_1^{\dagger}\Phi_1\Phi_2^{\dagger}\Phi_2+\lambda_{4}\Phi_1^{\dagger}\Phi_2 \Phi_2^{\dagger}\Phi_1+\left[\frac{ \lambda_5}{2}\left( \Phi_1^{\dagger}\Phi_2\right)^2+\text{h.c.}\right], \label{eq:potential}
 \end{align}
where the coefficients $\mu_{1}^2$, $\mu_{2}^2$, $\lambda_1,\lambda_2,\lambda_3,$ and $\lambda_4$, are real by hermiticity of the potential~\cite{Ginzburg:2003fe,Ginzburg:2004vp}, 
 whereas $\mu_{12}^2$ and $\lambda_5$ are assumed to be real\footnote{There are another quartic couplings, not considered here, which are forbidden if one of the Higgs doublets is odd under an extra $\mathbb{Z}_2$ symmetry softly violated by the $\mu_{12}^2$ bilinear term.}. To prevent the emergence of spontaneous $CP$ violation, we choose real and positive values for $v_1$ and $v_2$~\cite{Gunion:2002zf} and define
 \begin{align}
 \label{eq:tanb}
   \tan\beta=\frac{v_2}{v_1}\,.
 \end{align}
The quartic couplings $\lambda_1(\Lambda),...,\lambda_4(\Lambda)$ become constrained by the spectral action at the unification scale $\Lambda$, while $\mu_{12}^2$ and $\lambda_5(\Lambda)$ together with $\tan \beta$ are the free parameters of the model. 
 
The main result of our work is that the type-II 2HDM in NCG with the scalar potential in Eq.~\eqref{eq:potential},
accounts for the correct SM-like Higgs boson mass of $125$ GeV, involving extra
scalars compatible  with the alignment limit
without decoupling from masses around $350\ \text{GeV}$, for a very constrained
parameter space encompassing $\tan\beta\approx 2.1$.

The rest of the paper is organized as follows.
In Sec.~\ref{sec:NCG}, we introduce the main notions in NCG  to emphasize its underlying 2HDM structure. 
In Sec.~\ref{sec:spec}, we deduce the spectral action constrains of the model in order to get the boundary conditions for the quartic couplings of the 2HDM scalar potential.  
In Sec.~\ref{sec:mass}, we review the mass formulas for the scalar fields in 2HDM in preparation for the phenomenological analysis.
In Sec.~\ref{sec:pheno}, we scan the parameter space at the unification scale and evolve all the relevant Yukawa couplings and quartic scalar parameters down to the electroweak scale.
There, we check if there are regions of the parameter space compatible with a SM-like Higgs boson. 
Finally, we close with the conclusions of our work.

\section{Noncommutative geometry: Higgs sector}
\label{sec:NCG}

In this section, we introduce the main NCG tools which are necessary to  describe the SM emphasizing their underlying 2HDM structure.

The study of noncommutative geometry has its roots in the Gelfand-Naimark theorem, which establishes an equivalence between compact topological spaces and commutative,
C$^*$-algebras~\cite{Landi:1997sh,Gracia-Bondia:2001upu}. In consequence, the topology of a manifold $M$ is encoded by the algebra of (complex) continuously differentiable functions defined over the manifold $C^{\infty}(M,\mathbb{C})$, so that the points $x\in M$ correspond to irreducible representations of the algebra. 

In order to characterize the geometrical information of a Riemannian manifold, we introduce the spectral triple formulation (for complete reviews see e.g.~\cite{vandenDungen:2012ky,Farnsworth:2015dcv}). Let us suppose that $\mathcal{S}\rightarrow M$ is a spinor bundle on a 4-dimensional manifold $M$, so that spinor fields on $M$ are given by smooth sections $\psi$ in the Hilbert space of square-integrable sections $L^2(M,\mathcal{S})$. Then, we can represent the algebra elements by bounded operators acting on spinors as $(f\psi)(x)=f(x)\psi(x)$. The so-called `canonical spectral triple' is defined by $ \{C^{\infty}(M,\mathbb{C}),L^2(M,\mathcal{S}),\slashed D,\gamma_5, J_M\}$,
where the  Dirac operator is defined by
\begin{align*}
  \slashed D=-i\gamma^{\mu}\nabla^{\mathcal{S}}_{\mu}\,,
\end{align*}
with $\nabla^{\mathcal{S}}_\mu$ being the spin Levi-Civita connection on $\mathcal{S}$ and $\gamma^{\mu}$ are the usual gamma matrices which generate the Dirac algebra. Additionally,  the (Hermitian) chirality operator $\gamma_5$, and the anti-unitary charge conjugation $J_M$ are given after fixing a Dirac algebra basis
$\gamma^{\mu}$, such that 
\begin{align}
  \label{eq:rel1}
 \left[f,J_Mg^*J_M^{-1}\right]&=0,& \gamma_5^2=&1_4,&J_M^2&=-1_{4}, &J_M\gamma_5 =&\gamma_5 J_M\,, 
\end{align}
for all $f, g\in C^{\infty}(M,\mathbb{C})$. The Dirac operator $\slashed{D}$ satisfies the Leibniz rule
\begin{align*}
\nabla^\mathcal{S}_{\mu}({f}\psi)={f}\nabla^{\mathcal{S}}_{\mu}(\psi)+\partial_{\mu}({f})\psi\,,
\end{align*}
for all $f\in C^{\infty}(M,\mathbb{C})$ and $\psi \in L^2(M,\mathcal{S})$. It also satisfies the following commutation properties
\begin{align}
 J_M\slashed D&=\slashed DJ_M, &\slashed D\gamma_5 =&-\gamma_5 \slashed D,&\left[\left[\slashed{D},f\right],J_Mg^*J_M^{-1}\right]=0\,, \label{eq:rel2}
\end{align}
for all $f, g\in C^{\infty}(M,\mathbb{C})$. Note that the one-forms are bounded, because they are given by the commutator $\left[\slashed{D},f\right]=-i\gamma^{\mu}(\partial_{\mu}f)$, which is necessary to recover the metric information from $M$~\cite{Iochum:1999ih,Connes:1996gi}.

The next step in NCG is to inquire about the geometry of a noncommutative $*$-algebra. In this case, the points are mimicked by the set of `states' of the algebra\footnote{The states of an algebra are defined as linear maps of the form $\pi_x:A\rightarrow \mathbb{C}$, which are positive $\pi_x(f^*f) \geq 0 \in \mathbb{R}$,
for all $f$ in $A$ and so that $\pi_x(\mathbf{1}_A) = 1$, for $\mathbf{1}_A$ the unit of $A$.}, which corresponds to its set of characters in the case that the algebra is commutative~\cite{Farnsworth:2020ozj}.
For physical motives, we concentrate in the finite-dimensional algebra $A_F=\mathbb{C}\oplus\mathbb{H}\oplus M_3(\mathbb{C})$, where the second and last summands stand for the quaternion algebra and the set of $3\times3$ complex matrices, respectively. We introduce the `finite spectral triple' $  \{A_F,H_F,D_F,\gamma_F,J_F\} $, where the algebra elements ($D_F, \gamma_F$ and $ J_F$ will be defined below) can be represented as matrices acting on the finite-dimensional 
 (complex) Hilbert space~\cite{vanSuijlekom:2015iaa}
\begin{align*}
  H_F=(H_l\oplus H_q\oplus H_{\overline{l}}\oplus H_{\overline{q}})^{\oplus3}\,,
\end{align*}
which accounts for the three SM generations (without right-handed neutrinos), where the spaces of leptons $H_{l}=\mathbb{C}^3$ and the quarks $H_{q}=\mathbb{C}^4\otimes\mathbb{C}^3$ are spanned by $\{e_R,\nu_L,e_L\}$ and $\{u_R,d_R,u_L,d_L\}$, whereas the anti-lepton $H_{\overline{l}}=\mathbb{C}^3$ and  the anti-quark $H_{\overline{q}}=\mathbb{C}^4\otimes\mathbb{C}^3$ spaces are generated by the basis $\{\overline{e_R},\overline{\nu_L},\overline{e_L}\}$ and $\{\overline{u_R},\overline{d_R},\overline{u_L},\overline{d_L}\}$, respectively.
The action of an element $a=(\lambda,q,m)\in A_F$ (with $q\in \mathbb{H}$) on the particle side of the Hilbert space is defined by 
 \begin{align}
   \label{eq:rep}
   \pi(a)|_{H_l}&=\begin{pmatrix} \overline{\lambda}&0&0\\
0&\alpha&\beta\\
 0&-\overline{\beta}&\overline{\alpha}
 \end{pmatrix},  &\text{and}&&
 \pi(a)|_{H_q}&=\begin{pmatrix}
 {\lambda}&0&0&0\\
 0&\overline{\lambda}&0&0\\
0&0&\alpha&{\beta}\\
 0&0&-\overline{\beta}&\overline{\alpha}
 \end{pmatrix}\otimes1_3\,,
 \end{align}
where $ m\in M_3(\mathbb{C})\  \text{and} \ \lambda, \alpha, \beta\in\mathbb{C}$, and the bar over them stands for their complex conjugated.  The action of $a$ on the anti-particles is so that $\pi(a) \overline{l}=\lambda 1_3 \overline{l}$ and $\pi(a)\overline{q}=(1_4\otimes m) \overline{q}$. For easy use, from now on we will drop the symbol $\pi$ to denote the representation of an algebra element.
 
Regarding the other parts of the finite spectral triple, the finite charge conjugation operator $J_F$ is defined so that it interchanges particles with their respective antiparticles, whereas the finite chirality operator $\gamma_F$ assigns the eigenvalue $+1$ to left-handed particles, and the eigenvalue $-1$ to the right-handed ones. 
Finally, the finite self-adjoint Dirac operator, $D_F$, involves the Yukawa matrices for leptons and quarks as follows 
\begin{align*}
  D_F=\begin{pmatrix}
  S&T^{\dagger}\\
  T&\overline{S}
  \end{pmatrix},
       \end{align*}
where $S$ and $T$ are operators acting on the SM particles, while $T^{\dagger}$ and $\overline{S}$ act on the anti-particles.  The action of $S$ on the SM fermions is given by
\begin{align}
\label{eq:S}
  S|_{H_l}&={\begin{pmatrix}
  0&0&Y_{l,1}^{\dagger}&Y_{l,2}^{\dagger}\\
   Y_{l,1}&0&0&0\\
   Y_{l,2}&0&0&0
   \end{pmatrix} },
&
S|_{H_q}&=\begin{pmatrix}
0&0&Y_{q,11}^{\dagger}&Y_{q,21}^{\dagger}\\
0&0&Y_{q,12}^{\dagger}&Y_{q,22}^{\dagger}\\
   Y_{q,11}&Y_{q,12}&0&0\\
   Y_{q,21}&Y_{q,22}&0&0
   \end{pmatrix}\otimes1_3\,, 
\end{align}
where $Y_{l,i}$ and $Y_{q,ij}$ are $3\times3$ complex matrices. On the other hand, $T$ contains some unwanted terms which are usually  removed by introducing the non-geometric massless photon condition~\cite{Boyle:2014wba,Chamseddine:2006ep} or by making use of the second order axiom~\cite{Farnsworth:2015dcv}
\begin{align}
  \label{eq:second}
 \left[\left[D_F,a\right],J_F\left[D_F,b\right]^*J_F^{-1}\right]=0\,,  
\end{align}
for all $a,b\in A_F$. In the absence of right-handed neutrinos, the condition in Eq.~\eqref{eq:second} reduces $T$ to $0$. One advantage (which we will exploit later) of using the second-order axiom instead of the massless photon condition is that none of the $Y_{l,i}$ or $Y_{q,ij}$ in the Eq.~\eqref{eq:S} is required to be zero.
 
 In general, given an arbitrary spectral triple $\{A,H,D,\gamma,J \}$,
 the last two  relations in Eq.~\eqref{eq:rel1}, together with the first one in Eq.~\eqref{eq:rel2} can be  shifted to 
 \begin{align*}
J^2&=\epsilon\mathbb{I},&
JD&=\epsilon'DJ, & J\gamma &=\epsilon''\gamma J\,,
\end{align*}
where $\mathbb{I}$ stands  for the identity operator on $H$, while the symbols $\epsilon , \epsilon',\epsilon'' \in \{1,-1\}$ define the KO-dimension\footnote{The name `KO' comes from the Bott periodicity theorems for real K-theory based on real vector bundles~\cite{atiyah2000ktheory}.} (or signature) modulo $8$ as shown in Table~\ref{tab:KO}.
We choose   $\epsilon=1$, $\epsilon'=1$ and $\epsilon''=-1$ so that it
 belongs to  a KO-dimension $6$ modulo $8$ (see Table~\ref{tab:KO}), which is necessary to preserve the correct fermion degrees of freedom~\cite{Lizzi:1996vr,Connes:2006qv,Barrett:2006qq}.

\begin{table}
\centering
\begin{tabular}{|l|r|r|r|r|r|r|r|r|} \hline

 & 0 & 1 & 2 & 3 & 4 & 5 & 6 & 7 \\\hline

$\epsilon$ & 1 & 1 & -1 & -1 & -1 & -1 & 1 & 1\\

$\epsilon'$ & 1 & -1 & 1 & 1 & 1 & -1 & 1 & 1 \\

$\epsilon''$ & 1 & & -1 & & 1 & & -1 & \\\hline
\end{tabular}
\caption{Module 8 KO-dimension.}
\label{tab:KO}
\end{table}

To have a complete picture involving all of the SM fields, 
we take the tensor product between the canonical spectral triple and the finite one, which defines the  so-called almost-commutative manifold. In this case, the resulting spectral triple is given by  $\mathcal{A}=C^{\infty}(M,\mathbb{C})\otimes A_F$ represented on the tensor product space $\mathcal{H}=L^2(\mathcal{S},M)\otimes H_F$. The chirality and the charge conjugation operators are
\begin{align*}
  \Gamma=&\gamma_5\otimes \gamma_F\,,& \text{ and }&& \mathcal{J}=&J_M\otimes J_F\,,
\end{align*}
respectively, whereas the total Dirac operator is defined by
\begin{align}
  \label{eq:dirac}
  \mathcal{D}= \slashed D\otimes 1_{F}+\gamma_5\otimes D_F\,,
\end{align}
where $1_F$ stands for the identity matrix on the finite space $H_F$.

The fermion fields $\Psi\in\mathcal{H}$ should satisfy  
$ \Gamma\Psi=\Psi$~\cite{Connes:2006qv}, and are given by the tensor product between anti-commuting Weyl spinors and finite vector basis elements, both denoted by the same symbol as follows
\begin{align}
  \Psi\equiv &  e_L\otimes e_L +e_R\otimes e_R +\overline{e_L}\otimes \overline{e_L} +\overline{e_R}\otimes \overline{e_R}+\nu_L\otimes \nu_L +\overline{\nu_L}\otimes \overline{\nu_L}+u_L\otimes u_L \nonumber\\
  & +u_R\otimes u_R +\overline{u_L}\otimes \overline{u_L} +\overline{u_R}\otimes \overline{u_R}+ d_L\otimes d_L +d_R\otimes d_R +\overline{d_L}\otimes \overline{d_L} +\overline{d_R} \otimes\overline{d_R}\,,
\end{align}
where for simplicity we omitted family  and color indices.

The bosonic fields come along with the inner fluctuations of the Dirac operator, which are given by linear combinations of terms of the form $a[\mathcal{D},b]$, for  
$a, b \in \mathcal{A}$, which we write as $a=(\lambda,q,m)\,$ and $b=(\lambda',q',m')\,$~\cite{vanSuijlekom:2015iaa,vandenDungen:2012ky}. By using Eq.~eqref{eq:dirac}, we can split such commutator into the following two terms
\begin{align*}
  a[\slashed{D}\otimes1_F,b]&\equiv\gamma^{\mu}\otimes A_{\mu} \ ,& a[\gamma_5\otimes D_F,b]&\equiv\gamma_5\otimes \phi \,,
\end{align*}
where $A_{\mu}\equiv-ia\partial_{\mu}b\,$ and $\phi\equiv a[D_F,b]$ determine the Hermitian gauge and scalar fields, respectively. 

The gauge fields can be expressed as follows
\begin{align*}
A_{\mu}|_{H_l}&=\begin{pmatrix}
\Lambda_{\mu}&0\\
0&Q_{\mu}
\end{pmatrix},
&
A_{\mu}|_{H_{\overline{l}}}&=\Lambda_{\mu} 1_3\,,\\
A_{\mu}|_{H_q}&=\begin{pmatrix}
\Lambda_{\mu}&0&0\\
0&-\Lambda_{\mu}&0\\
0&0&Q_{\mu}
\end{pmatrix}\otimes1_3,
&
A_{\mu}|_{H_{\overline{q}}}&=-\left(\overline{V}_{\mu}+\frac{1}{3}\Lambda_{\mu}\right)\otimes1_4\,,
\end{align*}
where we have defined
\begin{align*}
   \Lambda_{\mu}\equiv& -i\overline{\lambda}\partial_{\mu}\overline{\lambda}'\,, &Q_{\mu}\equiv& -iq\partial_{\mu}q'\,, &V_{\mu}\equiv& -im\partial_{\mu}m'\,.
\end{align*}
They can be compassed into the total gauge field $B_{\mu}\equiv A_{\mu}-J_FA_{\mu}J_F^{\dagger}$, whose action on the particle side is given by
 \begin{align*}
B_{\mu}|_{H_l\oplus H_q}=\operatorname{diag}\left[-2\Lambda_{\mu},Q_{\mu}-\Lambda_{\mu}1_2,V_{\mu}+\frac{4}{3}\Lambda_{\mu}\otimes1_3,V_{\mu}-\frac{2}{3}\Lambda_{\mu}\otimes1_3,V_{\mu}\otimes1_2+\left(Q_{\mu}+\frac{1}{3}\Lambda_{\mu}1_2\right)\otimes1_3\right].
\end{align*}
Note that the coefficients of $\Lambda_{\mu}$ are the hypercharges of the corresponding SM fermions.

On the other hand, with the hermiticity of $\phi$, we have that the action of the Higgs field,
\begin{align*}
  {\Phi}\equiv D_F+ \phi+J_F \phi J_F^{\dagger}\,,
\end{align*}
on the particles, is given by 
a general 2HDM Yukawa structure with four complex degrees of freedoms
\begin{align}
  \Phi|_{H_l}&=\begin{pmatrix}
 0&\overline{\varphi}_1Y_{l,1}^{\dagger}+\overline{\varphi}_2Y_{l,2}^{\dagger}& \overline{\varphi}_3Y_{l,1}^{\dagger}+\overline{\varphi}_4Y_{l,2}^{\dagger}\\
\varphi_1Y_{l,1}+\varphi_2Y_{l,2}&0&0\\
 \varphi_3Y_{l,1}+\varphi_4Y_{l,2}&0&0
  \end{pmatrix},\nonumber\\
  \Phi|_{H_q}&=\begin{pmatrix}
 0&0&{\varphi}_4Y_{q,11}^{\dagger}-{\varphi}_3Y_{q,21}^{\dagger}& -{\varphi}_2Y_{q,11}^{\dagger}+{\varphi}_1Y_{q,21}^{\dagger}\\
 0 &0&\overline{\varphi}_1Y_{q,12}^{\dagger}+\overline{\varphi}_2Y_{q,22}^{\dagger}&\overline{\varphi}_3Y_{q,12}^{\dagger}+\overline{\varphi}_4Y_{q,22}^{\dagger}\\
\overline{\varphi}_4Y_{q,11}-\overline{\varphi}_3Y_{q,21}&\varphi_1Y_{q,12}+\varphi_2Y_{q,22}&0&0\\
 -\overline{\varphi}_2Y_{q,11}+\overline{\varphi}_1Y_{q,21}&\varphi_3Y_{q,12}+\varphi_4Y_{q,22}&0&0
  \end{pmatrix},
\end{align}
where, using Eqs.~\eqref{eq:rep}, we define
\begin{align*}
\varphi_1&\equiv \alpha(\overline{\lambda}'-{\alpha'})+\beta\overline{\beta}'+1\,,&\varphi_2&\equiv \beta(\overline{\lambda}'-\overline{\alpha}')-\alpha\beta'
\,,\\
\varphi_3&\equiv \overline{\beta}(\alpha'-\overline{\lambda}')+\overline{\alpha}\overline{\beta}'\,,& \varphi_4&\equiv \overline{\alpha}(\overline{\lambda}'-\overline{\alpha}')+\overline{\beta}\beta'+1\,.
\end{align*}
We observe that with the identification 
\begin{align}
\label{eq:Phi12}
\Phi_1=&\begin{pmatrix}
\varphi_1\\
\varphi_3
\end{pmatrix},&
\Phi_2=&\begin{pmatrix}
\varphi_2\\
\varphi_4
\end{pmatrix},
\end{align}
the Higgs sector in NCG is in general composed of two complex Higgs doublets, as mentioned in~\cite{Farnsworth:2014vva}.
To avoid the appearance of flavor changing neutral currents (FCNC) at tree-level, we use the usual approach of impose a (softly broken~\cite{Bernon:2015qea,Ginzburg:2004vp})
$\mathbb{Z}_2$ type-II 2HDM symmetry~\cite{Branco:2011iw} enforced with $\Phi_1 \rightarrow -\Phi_1$,  $d_R\to -d_R$, and $e_R\to -e_R$. This implies that  $Y_{l,2}=Y_{q,21}=Y_{q,22}=0$, so that    
 by defining $Y_u\equiv Y_{q,11}$, $Y_d\equiv Y_{q,12}$ and $Y_e\equiv Y_{l,1}$, we can express the Yukawa interaction as\footnote{This is a piece of the total fermionic action which is given by the bilinear form  $\langle J\Psi,\mathcal{D}_A\Psi\rangle$, where $\mathcal{D}_A=\mathcal{D}+a[\mathcal{D},b]+\mathcal{J}a[\mathcal{D},b]\mathcal{J}^{\dagger}$ is the fluctuation of the total Dirac operator~\cite{Connes:2006qv}.}
 \begin{align}
\label{eq:Yuk}   
\frac{1}{2} \langle J\Psi,\left( \gamma_5\otimes{\Phi}\right)\Psi\rangle&= \langle J_M\overline{u_R}, Y_u^{\dagger} \left({\varphi}_{4} {u} _L-{\varphi}_{2}{d} _L\right)\rangle  +\langle J_ M \overline{d_R},Y_d^{\dagger}\left(\overline{\varphi}_{1}{u} _L+\overline{\varphi}_{3}{d} _L\right) \rangle\nonumber\\
&+\langle J_ M{\overline{e_R}}, Y_e^{\dagger} \left(\overline{\varphi}_{1}{\nu}_L+\overline{\varphi}_{3} {e}_L\right)\rangle
-\text{h.c}\,,
\end{align}
where the minus sign preceding the `h.c.' acronym means that the Yukawa matrices are anti-Hermitian. Note that   the Lagrangian in Eq.~\eqref{eq:Yuk} corresponds to the type-II 2HDM Yukawa interaction~\cite{Hall:1981bc,Atkinson:2021eox}.

Now, we examine the spectral action principle, from which we will deduce the constraints on the parameter space of the model at the unification scale.

\section{ Spectral action and constrains }
\label{sec:spec}

The action functional describing the dynamics of the SM bosons is given by the spectral action
 $\operatorname{Tr}\left(f(\mathcal{D}_{A}/\Lambda)\right)$, where $\Lambda$ is a cut-off scale, and $f$ is a real, even, and positive function such that $f\rightarrow0$ when $\Lambda \rightarrow \infty$. By writing $f$ in terms of its Laplace transform, we can use the heat kernel expansion~\cite{Vassilevich:2003xt} to expand the spectral action, so that we concentrate only in the following piece~\cite{Iochum:2012bu, Fan:2015hva} 
\begin{align}
\label{eq:sectact}   
\operatorname{Tr}\left[f\left(\frac{D_{A}}{\Lambda}\right)\right]\supset \int_M\frac{f(0)}{8\pi^2}\left\{\frac{1}{3}\operatorname{Tr}\left(F^{\mu\nu}F_{\mu\nu}\right)+\operatorname{Tr}\left[\left(\widetilde{D}_{\mu}\Phi\right)
\left(\widetilde{D}^{\mu}\Phi\right)
\right]+\operatorname{Tr}\left({\Phi}^4\right)\right\}\sqrt{|g|}\operatorname{d}^4x\,,
\end{align}

where $F^{\mu\nu}$ is the curvature of $B_{\mu}$, and the second term is given by $
\widetilde{D}_{\mu}{\Phi}=\partial_{\mu}\Phi+i[B_{\mu},\Phi]
$.

Having into account the three SM generations of fermions and anti-fermions, the first term in Eq.~\eqref{eq:sectact} is given by 
 \begin{align*}
\operatorname{Tr}\left(F^{\mu\nu}F_{\mu\nu}\right)&=24\left[\frac{10}{3}\operatorname{Tr}\left(\Lambda_{\mu\nu}\Lambda^{\mu\nu}\right)+\operatorname{Tr}\left(Q_{\mu\nu}Q^{\mu\nu}\right)+\operatorname{Tr}\left(V_{\mu\nu}V^{\mu\nu}\right)\right],
\end{align*}
where $\Lambda_{\mu\nu}$, $Q_{\mu\nu}$ and $V_{\mu\nu}$ are the curvature of $\Lambda_{\mu}$, $Q_{\mu}$ and $V_{\mu}$, respectively.
We introduce the gauge couplings $g_1$, $g_2$ and $g_3$ by rescaling the gauge fields in terms of the SM vector fields $Y_{\mu}$, $W_{\mu}$ and $G_{\mu}$ as 
\begin{align*}
 \Lambda_{\mu}\equiv& \frac{g_1}{2}Y_{\mu}\,,& Q^i_{\mu}\equiv& \frac{g_2}{2}W^i_{\mu}\,,& V^i_{\mu}\equiv& \frac{g_3}{2}G^a_{\mu}\,,
\end{align*}
with $i=1,2,3$ and $a=1,2,\ldots, 8\,$.
The gauge couplings can be properly normalized so that~\cite{vanSuijlekom:2015iaa}
\begin{align}
  \label{eq:fcero}  
  \frac{g_2^2f(0)}{2\pi^2}&=\frac{1}{4}\,,&g\equiv \frac{5}{3}g_1^2&=g_2^2=g_3^2\,,
\end{align}
where we have introduced $g$ as an unified gauge coupling constant. Since the constrain on the gauge couplings, established by Eq.~\eqref{eq:fcero}, is the  same that holds for grand unified theories, 
we assume that the cut-off scale, $\Lambda$, of the spectral action is the grand unification scale.

The last term in Eq.~\eqref{eq:sectact} is given by
\begin{align}
  \label{eq:first}  
  \operatorname{Tr}\left({\Phi}^4\right)=& 4\left[\operatorname{Tr}\left(Y_e^{\dagger}Y_e\right)^2+3\operatorname{Tr}\left(Y_d^{\dagger}Y_d\right)^2\right]\left(\Phi^{\dagger}_1\Phi_1\right)^2+12\operatorname{Tr}\left(Y_u^{\dagger}Y_u\right)^2\left(\Phi^{\dagger}_2\Phi_2\right)^2\nonumber\\
                 &+12 \left[\operatorname{Tr}\left(Y_u^{\dagger}Y_uY_d^{\dagger}Y_d\right)+\operatorname{Tr}\left(Y_uY_u^{\dagger}Y_dY_d^{\dagger}\right)\right]\left[\left(\Phi^{\dagger}_1 \Phi_1\right)\left(\Phi^{\dagger}_2 \Phi_2\right)-\left(\Phi^{\dagger}_1 \Phi_2\right)\left(\Phi^{\dagger}_2 \Phi_1\right)\right].
\end{align}
To get the  
kinetic terms for the scalar fields, we first calculate the commutator $[B_{\mu},\Phi]$, from which we obtain
\begin{align}
\frac{f(0)}{8\pi^2}\operatorname{Tr}\left[\left(\widetilde{D}_{\mu}\Phi\right)\left(   \widetilde{D}^{\mu}\Phi\right)\right]&=\frac{\operatorname{Tr}\left(Y_e^{\dagger}Y_e\right)+3\operatorname{Tr}\left(Y_d^{\dagger}Y_d\right)}{4g^2}\left(D_{\mu}\Phi_1\right)^{\dagger}\left(D_{\mu}\Phi_1\right)+\frac{3\operatorname{Tr}\left(Y_u^{\dagger}Y_u\right)}{4g^2}\left(D_{\mu}\Phi_{2}\right)^{\dagger}\left(D_{\mu}\Phi_{2}\right),
 \end{align}
 where $D_{\mu}\Phi_{j}=\partial_{\mu} \Phi_{j} +iQ_{\mu}^i\sigma_i\Phi_{j}+i\Lambda_{\mu} \Phi_{j}$, with $j=1,2$, stands for the covariant derivative acting on the Higgs doublets.  
To have a proper normalization for this equation, we redefine the scalar doublets as 
\begin{align}
\label{eq:norm11}  
\Phi_1&\to \frac{2g}{\sqrt{\operatorname{Tr}\left(Y_e^{\dagger}Y_e\right)+3\operatorname{Tr}\left(Y_d^{\dagger}Y_d\right)}}\Phi_{1},&
{\Phi}_{2}&\to  \frac{2g}{\sqrt{3\operatorname{Tr}\left(Y_u^{\dagger}Y_u\right)}}{\Phi}_2\,,
  \end{align}
and from now on 
\begin{align}
\label{eq:fHD}
\Phi_2=&\begin{pmatrix}
\phi_2^+\\
\phi_2^0\\
\end{pmatrix},&\Phi_{1}=&\begin{pmatrix}
\phi_1^+\\ \phi_1^0
\end{pmatrix}.
\end{align}
In terms of these normalized fields, we can re-write Eq.~\eqref{eq:first} to get the quartic scalar terms as follows
\begin{align*}
 \frac{f(0)}{8\pi^2}\operatorname{Tr}({\Phi}^4) &=4g^2 \frac{\operatorname{Tr}\left(Y_e^{\dagger}Y_e\right)^2+3\operatorname{Tr}\left(Y_d^{\dagger}Y_d\right)^2}{\left[\operatorname{Tr}\left(Y_e^{\dagger}Y_e\right)+3\operatorname{Tr}\left(Y_d^{\dagger}Y_d\right)\right]^2}\left(\Phi_{1}^{\dagger}\Phi_{1}\right)^2+\frac{4}{3}g^2\left(\Phi_{2}^{\dagger}\Phi_{2}\right)^2\nonumber\\
 &+4g^2\frac{\operatorname{Tr}\left(Y_u^{\dagger}Y_uY_d^{\dagger}Y_d\right)+\operatorname{Tr}\left(Y_uY_u^{\dagger}Y_dY_d^{\dagger}\right)}{\operatorname{Tr}\left(Y_u^{\dagger}Y_u\right)\left[\operatorname{Tr}\left(Y_e^{\dagger}Y_e\right)+3\operatorname{Tr}\left(Y_d^{\dagger}Y_d\right)\right]} \left[\left(\Phi_1^{\dagger}\Phi_1\right)\right(\Phi_{2}^{\dagger}\Phi_{2}\left)-\left(\Phi_1^{\dagger}\Phi_{2}\right)\left(\Phi_1^{\dagger}\Phi_{2}\right)\right].
\end{align*}
By comparing this expression with the quartic scalar potential terms in Eq~\eqref{eq:potential}, we make the following identification
\begin{subequations}
\begin{align}
{\lambda_2(\Lambda)}&=\frac{8}{3}g^2\,,\\
\lambda_1(\Lambda)&=8g^2 \frac{\operatorname{Tr}\left(Y_e^{\dagger}Y_e\right)^2+3\operatorname{Tr}\left(Y_d^{\dagger}Y_d\right)^2}{\left[\operatorname{Tr}\left(Y_e^{\dagger}Y_e\right)+3\operatorname{Tr}\left(Y_d^{\dagger}Y_d\right)\right]^2}\,,\\\lambda_3(\Lambda)&=-\lambda_4(\Lambda)=4g^2\frac{\operatorname{Tr}\left(Y_u^{\dagger}Y_uY_d^{\dagger}Y_d\right)+\operatorname{Tr}\left(Y_uY_u^{\dagger}Y_dY_d^{\dagger}\right)}{\operatorname{Tr}\left(Y_u^{\dagger}Y_u\right)\left[\operatorname{Tr}\left(Y_e^{\dagger}Y_e\right)+3\operatorname{Tr}\left(Y_d^{\dagger}Y_d\right)\right]}\,,
\end{align}
\label{eq:boundary}
\end{subequations}
which, in turn, constrain the values of the quartic scalar couplings at the unification scale $\Lambda$. 

From our normalization in Eq.~\eqref{eq:norm11}, we find that the (anti-Hermitian) Yukawa matrix for the quark-up in Eq.~\eqref{eq:Yuk} is given by
\begin{align}
  \label{eq:ytop}
  Y_u&\equiv -i\frac{\sqrt{3\operatorname{Tr}(Y_u^{\dagger}Y_u)}}{\sqrt{2}gv_2}m_u\,,
\end{align}
where $m_u$, is a Hermitian mass matrix.

From now on, we concentrate only on the third SM generation, so that the indices $u$, $d$, $e$ correspond to top ($t$), bottom ($b$), and tau ($\tau$), respectively. Therefore, we can rewrite Eqs.~\eqref{eq:boundary} as 
\begin{align}
\label{eq:boundary3}
\lambda_1(\Lambda)&=\frac{8(1+3\rho^2)}{(1+3\rho)^2}g^2, &
\lambda_2(\Lambda)&=\frac{8}{3}g^2,&\lambda_3(\Lambda)&=\frac{8\rho}{1+3\rho}g^2,&\lambda_4(\Lambda)&=-\frac{8\rho}{1+3\rho}g^2\,, 
\end{align}
where we have introduced the parameter 
\begin{align}
\label{eq:rho}
\rho=\frac{|y_b|^2}{|y_{\tau}|^2}\,.  
\end{align}
Analogously, by multiplying Eq.~\eqref{eq:ytop} times its own Hermitian conjugate, we obtain
\begin{align}
\label{eq:BY}
  y_t(\Lambda)=\frac{2g}{\sqrt{3}}\,.
\end{align}
Eqs.~\eqref{eq:boundary3} and~\eqref{eq:BY} are interpreted as the boundary conditions imposed by the spectral action on the quartic and the Yukawa couplings at the grand unification scale, respectively.
However, the coefficients  $\mu_{12}^2$ and $\lambda_5(\Lambda)$ are still unconstrained by the spectral action, and we consider them as the free parameters of the model.
Although the terms $\mu_1^2\Phi_1^{\dagger}\Phi_1+\mu_2^2\Phi_2^{\dagger}\Phi_2$ are in general present in the spectral action expansion (which could have cosmological implications~\cite{Chamseddine:2006ep, Stephan:2009te}), their values  are determined after symmetry breaking by the tadpole equations described in the next section.

\section{Scalar mass eigenstates}
\label{sec:mass}

Before going through the calculation of the mass terms of the scalar fields of the model, we present the additional constraints on the parameters $\lambda_i$ which guarantee the boundedness from below~\cite{Ivanov:2006yq, Ferreira:2009jb} of the potential in Eq.~\eqref{eq:potential}
\begin{subequations}
\begin{align}
  \lambda_1&>0\,, & \lambda_3&>-\sqrt{\lambda_1\lambda_2}\,, \\
  \lambda_2&>0\,, & \lambda_3+\lambda_4-|\lambda_5|&>-\sqrt{\lambda_1\lambda_2}\,.
\end{align}
\label{eq:bound0}
\end{subequations}

In the two-Higgs-doublet model, after electroweak symmetry breaking, a total of eight scalar degrees of freedom appear. Three of them are the massless Goldstone modes $G^\pm$ and $G^0$, which are absorbed to get the masses for SM gauge bosons $W^\pm$ and $Z$. The remaining fields are two $CP$-even (real-uncharged) scalars $h$, and $H$, one $CP$-odd pseudoscalar $A$, and two charged Higgs bosons $H^\pm\,$. 

By expanding the two Higgs doublets in Eq.~\eqref{eq:fHD} around the VEVs we get
\begin{align*}
\Phi_j=\begin{pmatrix}\phi_j^+\\
\frac{1}{\sqrt{2}}\left(h_j+v_j+i\eta_j\right)
 \end{pmatrix},\hspace{0.35cm}\text{for}\hspace{0.25cm}j=1,2\,,
    \end{align*}
where $\Re(\phi^0_j)=h_j+v_j$ and $\Im(\phi^0_j)=\eta_j$. The    ratio between the two VEVs, $\tan\beta$ in Eq.~\eqref{eq:tanb}, is defined for $0\leq \beta <{\pi}/{2}$, with $v=\sqrt{v_1^2+v_2^2}=246.2$ GeV.
  The minimum conditions ${\partial V/\partial \Phi_j}|_{\langle\Phi_j\rangle_0}=0$ for the potential give rise to
\begin{align*}
  \mu_1^2&=\mu_{12}^2\tan{\beta}-\frac{\lambda_1}{2}v_1^2-\frac{\lambda_{345}}{2}v_2^2, \nonumber\\ \mu_2^2&=\mu_{12}^2\cot{\beta}-\frac{\lambda_2}{2}v_2^2-\frac{\lambda_{345}}{2}v_1^2\,, 
\end{align*}
where $\lambda_{345}\equiv \lambda_{3}+\lambda_4+\lambda_5 $.

 The physical scalar states are obtained by means of the following rotation matrices 
\begin{align*}
  \begin{pmatrix}
H\\
h
\end{pmatrix}&= {R_{\alpha}} \begin{pmatrix}
h_1\\
h_2
\end{pmatrix},&
 \begin{pmatrix}
G^{+}\\
H^{+}
\end{pmatrix}&={R_{\beta}} \begin{pmatrix}
\phi_1^+\\
\phi_2^+
\end{pmatrix},&\begin{pmatrix}
G^{0}\\
A
\end{pmatrix}&= {R_{\beta}} \begin{pmatrix}
\eta_1\\
\eta_2
\end{pmatrix},
\end{align*}
which are given in terms of the mixing angles $\alpha$ and $\beta$
\begin{align*}
  {R_{\alpha}}&= \begin{pmatrix}
\cos\alpha&\sin\alpha\\
-\sin\alpha&\cos\alpha
\end{pmatrix}, &
{R_{\beta}}&=\begin{pmatrix}
\cos\beta&\sin\beta\\
-\sin\beta&\cos\beta
\end{pmatrix}.
  \end{align*}
 The angle $\alpha$ is given by
 \begin{align}
   \tan 2\alpha=\frac{2\mathcal{M}^2_{12}}{\mathcal{M}^2_{11}-\mathcal{M}^2_{2 2}}\,,
 \end{align}
  where $-\pi/2<\alpha<\pi/2$ and $\mathcal{M}_{ij}^2$ are the components of the real uncharged square-mass matrix 
 \begin{align*}
  \mathcal{M}^2=\begin{pmatrix}\phantom{-}\mu_{12}^2\tan{\beta}+\lambda_{1}v_1^2&-\mu_{12}^2+\lambda_{345}v_1v_2\\
-\mu_{12}^2+\lambda_{345}v_1v_2&\phantom{-}\mu_{12}^2\cot{\beta}+\lambda_{2}v_2^2
\end{pmatrix}.
\end{align*}
 The mass expressions for the scalar fields are
 \begin{subequations}
 \label{eq:Mas}
     \begin{align}
  m^2_{H,h}&=\frac{1}{2}\left[\mathcal{M}_{11}^2+\mathcal{M}_{22}^2\pm\sqrt{\left(\mathcal{M}_{11}^2-\mathcal{M}_{22}^2\right)^2+4\left(\mathcal{M}_{12}^2\right)^2}\right],\\
m^2_{H^{\pm}}&=\left(\frac{\mu_{12}^2}{v_1v_2}-\frac{\lambda_{4}+\lambda_{5}}{2}\right)v^2,\\
m^2_{A}&=\left(\frac{\mu_{12}^2}{v_1v_2}-\lambda_{5}\right)v^2\,.
\end{align}
\end{subequations}
 We identify the $CP$-even state $h$ with the SM-like Higgs boson, whereas the other is the heavy $CP$-odd real scalar not yet discovered at the LHC. The main bounds in the scalar mass spectrum in Eqs.~ \eqref{eq:Mas} arises from direct Higgs searches and the coupling of the SM-like Higgs boson $h$ to the gauge bosons $W^{\pm}$ and $Z$. In the 2HDM, the coupling associated with this vertex is affected by the factor $\sin(\beta-\alpha)$. 
The SM couplings are recovered in the alignment limit~\cite{Bernon:2015qea} $\sin(\beta-\alpha)\approx1$, so that the vector bosons acquire their masses mainly by interacting with the observed 125 GeV $CP$-even state. 
We will make use of these restrictions in the next section, where we perform a RGEs analysis in order to derive the 2HDM mass spectrum in NCG.

\section{Phenomenology} 
\label{sec:pheno}
We fix the boundary conditions at $\Lambda=10^{16}\ \text{GeV}$ for the RGEs of the 2HDM couplings in Eq.~\eqref{eq:boundary3} and the top Yukawa in Eq.~\eqref{eq:BY}, and use the typical unification gauge coupling value $g=0.53$. 

We start by evolving the RGEs for~\cite{Basler:2017nzu}
\begin{align*}
16 \pi^2 \beta_{y_b}&=y_b\left(-8g_3^2-\frac{9}{4}g_2^2-\frac{5}{12} g_1^2+\frac{9}{2}y_b^2+\frac{1}{2}y_t^2+y_{\tau}^2\right),\\
16 \pi^2 \beta_{y_{\tau}}&=y_{\tau}\left(-\frac{9}{4}g_2^2-\frac{15}{4} g_1^2+\frac{5}{2}y_{\tau}^2+3y_b^2\right),
\end{align*}
from the electroweak scale, $m_Z$
\begin{align}
  y_{b}(m_Z)&=\frac{2.85 \sqrt{2}}{v\cos\beta}\,, &y_{\tau}(m_Z)&=\frac{1.78 \sqrt{2}}{v\cos\beta}\,,
\end{align}
 to $\Lambda$. Then, we fix $y_t(\Lambda)$ according to Eq.~\eqref{eq:BY} and we obtain $\tan\beta$ after get the proper value of $y_t(m_Z)$ by using~\cite{Basler:2017nzu}
 \begin{align}
  16 \pi^2 \beta_{y_t}&=y_t\left(-8g_3^2-\frac{9}{4}g_2^2-\frac{17}{12} g_1^2+\frac{9}{2}y_t^2+\frac{1}{2}y_b^2\right).
 \end{align}
 The direct measurements on top quark mass $172.9 \pm 0.4$ GeV~\cite{ParticleDataGroup:2018ovx} at $3\sigma$
 restrict $2.0\lesssim \tan\beta\lesssim 2.2$.

 In the same way, we vary the free parameter $\lambda_5(\Lambda)$ and use the boundary conditions, $\lambda_i(\Lambda)$, in Eq.~\eqref{eq:boundary3} to obtain 
 the quartic scalar couplings of the scalar potential at $m_Z$ from (see appendix A of both refs.~\cite{Basler:2017nzu,Ferreira:2009jb})
 \begin{subequations}\begin{align}
16 \pi^2 \beta_{\lambda_ 1}&=12 \lambda_ 1^2 + 
 4 \lambda_ 3^2 + 4 \lambda_ 3 \lambda_ 4+ 2 \lambda_ 4^2+2\lambda_5^2\nonumber\\&+\frac{3}{4} g_ 1^4 + \frac{3}{2} g_ 1^2 g_ 2^2 +\frac{9}{4} g_ 2^4 - 
 3 \left(g_ 1^2 + 3 g_ 2^2 \right)\lambda_ 1\,,\\
16 \pi^2 \beta_{\lambda_ 2}&= 12 \lambda_2^2 + 4 \lambda_3^2 +
 4 \lambda_3 \lambda_4 + 2 \lambda_4^2+2\lambda_5^2\nonumber\\&+\frac{3}{4} g_1^4 + \frac{3}{2} g_1^2 g_2^2 + \frac{9}{4} g_2^4 - 
 3\left( g_1^2 + 3 g_2^2 -
 4 y_t^2\right) \lambda_2 - 12 y_t^4\,,\\
16 \pi^2 \beta_{\lambda_ 3}&= \left(6 \lambda_3 +2 \lambda_4 \right)\left( \lambda_1 + \lambda_2\right)+
4 \lambda_3^2 + 
 2 \lambda_4^2+2\lambda_5^2\nonumber\\&+ \frac{3}{4} g_1^4 - \frac{3}{2} g_1^2 g_2^2 + \frac{9}{4} g_2^4 - 
 3 \left(g_1^2 +3 g_2^2 -
 2 y_t^2 \right)\lambda_3\,,\\
16 \pi^2 \beta_{\lambda_ 4}&= 2 \lambda_1 \lambda_4 + 2 \lambda_2 \lambda_4 + 
 8 \lambda_3 \lambda_4 + 4 \lambda_4^2+8\lambda_5^2\nonumber\\&+3 g_1^2 g_2^2 - 3 \left(g_1^2 + 
 3g_2^2 - 2 y_t^2\right) \lambda_4\,, \\
 16 \pi^2 \beta_{\lambda_ 5}&= \left(2 \lambda_1 + 2 \lambda_2 + 
 8 \lambda_3 + 12 \lambda_4\right)\lambda_5- 3 \left(g_1^2 + 
 3g_2^2 - 2 y_t^2\right) \lambda_5\,.
\end{align}\label{eq:rgei}
\end{subequations}

At the electroweak scale, we fix the other free parameter $\mu_{12}^2$.

To determine the scalar mass spectrum of this model, we scan the parameter space according to the ranges shown in Table~\ref{tab:ps}. We have chosen $|\lambda_5(\Lambda)|<0.5$ to preserve the perturbativity of the potential,
and $\mu_{12}^2>10^4$ GeV$^2$ to get rid of taquionic masses and to have a sufficiently  charged Higgs mass compatible with LEP constraints~\cite{Kling:2018xud,Li:2020hao,Han:2013mga}.

\begin{table}
\centering
\begin{tabular}{|c|c
|}\hline
Parameter& Range \\\hline
$\operatorname{sgn}(\lambda_5(\Lambda))$&$\pm 1$\\\hline
$|\lambda_5(\Lambda)|$&$(10^{-3},0.5)$\\\hline
$\mu_{12}^2 $&$(10^4\ \text{GeV}^2 ,10^6\ \text{GeV}^2)$ \\\hline
$\tan\beta$&$(2.0,2.2)$\\\hline
\end{tabular}
\caption{Scan range of the free parameters. We fix $\Lambda=10^{16}$~GeV and $g=0.53$\,.}
\label{tab:ps}
\end{table}

\begin{figure}
\centering
 \includegraphics[scale=0.65  ]{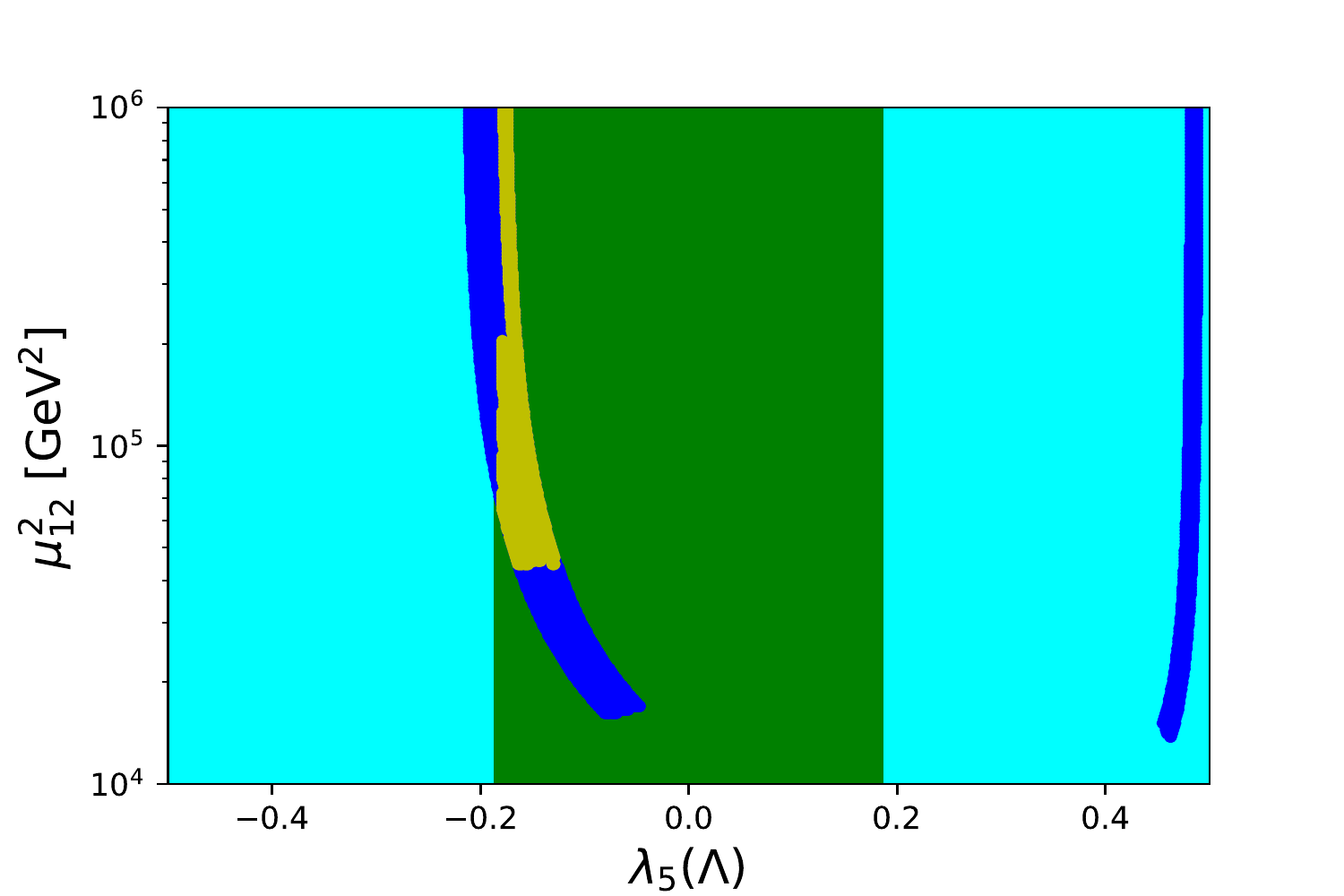}
\caption{ Scan of input parameters $\lambda_5(\Lambda)$ versus $\mu_{12}^2$. The green area corresponds to a potential which is bounded from below. The points within the yellow and dark-blue region satisfy the SM-like Higgs mass at $3\sigma$, whereas the dark blue region is excluded at 95\% C.L.}
\label{fig:mh}
\end{figure}

The explored parameter space for $\lambda_5(\Lambda)$ and $\mu_{12}^2$ is displayed as the cyan region in Fig.~\ref{fig:mh}.
After impose the theoretical constraints in Eqs.~\eqref{eq:bound0} in order to keep the boundedness from below of the potential, we get the green band in Fig.~\ref{fig:mh}. We use the implemented type-II 2HDM in SARAH~\cite{Staub:2008uz} to obtain the input file and the customized code SPheno~\cite{Porod:2003um,Porod:2011nf}, from which we can obtain the masses and mixing for each set of input parameters. The points compatible with a SM-like Higgs boson mass ($m_h=125.10 \pm 0.14 $ GeV~\cite{ParticleDataGroup:2018ovx}) at $3\sigma$, are shown as the yellow and dark-blue region in Fig.~\ref{fig:mh}. For each parameter point in those regions, we use the SPheno interface to HiggsBounds-5~\cite{Bechtle:2020pkv},
to test the neutral and charged Higgs sectors against the current exclusion bounds from the Higgs searches at the LEP, Tevatron and LHC experiments. With this tool, we determine that the parameter points in the dark-blue region are excluded at 95\% C.L. In this way, the remaining yellow region 
satisfy all the constraints, including the proper mass for the SM-like Higgs.

\begin{figure}
   \centering
   \includegraphics[scale=0.64]{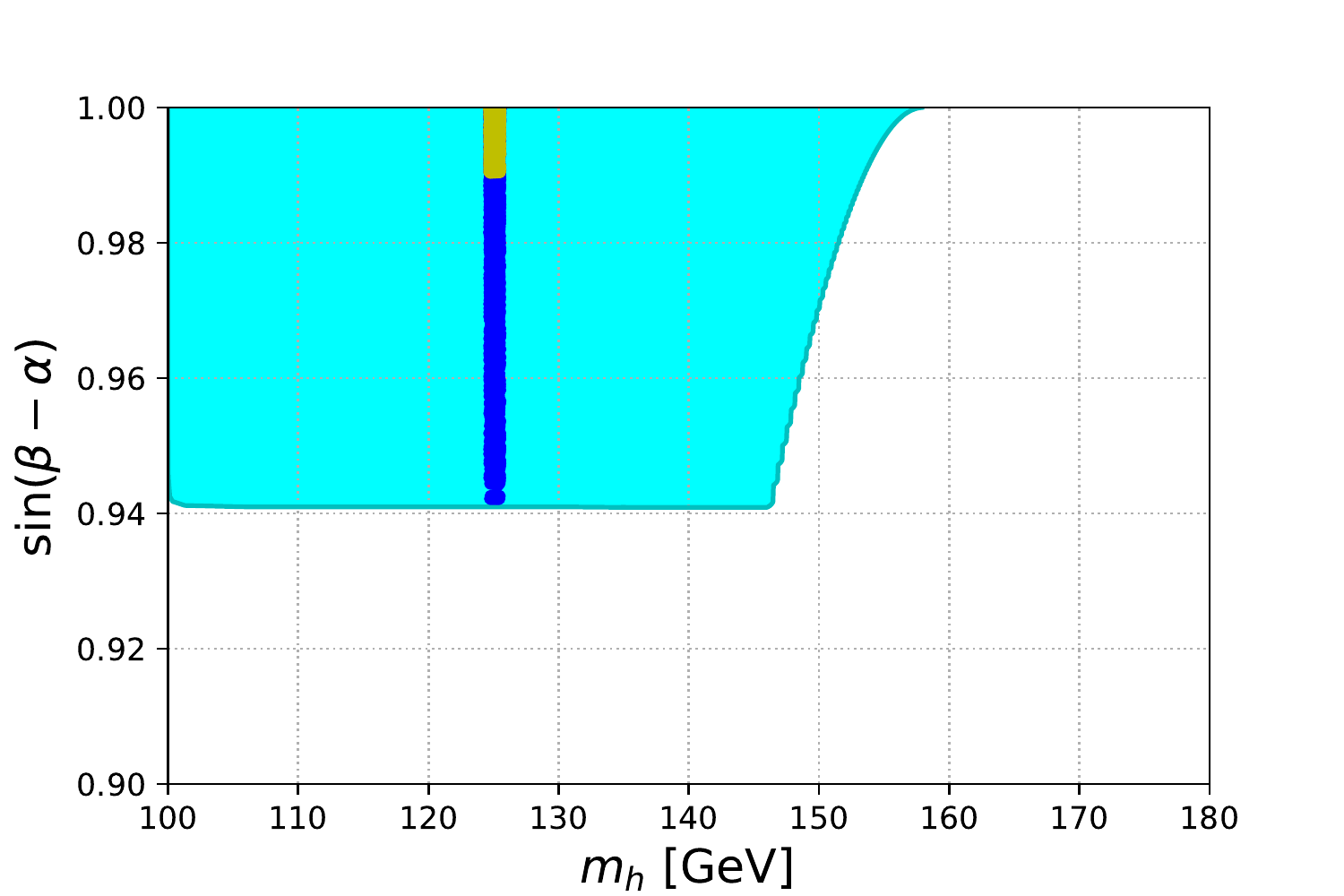}
   \caption{Scan of $m_h$ versus $\sin(\beta-\alpha)$. The 3$\sigma$ region for the SM-like Higgs mass corresponds to the yellow and dark blue region, while the yellow region satisfy all the current constraints as in Fig.~\ref{fig:mh}.}
   \label{fig:largemh}
 \end{figure}

The parameter space in Fig.~\ref{fig:mh} where the $m_h$ eigenvalue is less than $180\ \text{GeV}$ is shown in Fig.~\ref{fig:largemh}, where the dark-blue and yellow region corresponds to the very same points as in Fig.~\ref{fig:mh}, such that the yellow region can be identified with the SM-like Higgs compatible with all the current collider constraints\footnote{Our general scan includes another region with $m_h>185\ \text{GeV}$ and $\sin(\beta-\alpha)<0.6$ which is excluded by the LHC.}. In our approach, it is no possible to recover exactly the SM Higgs scalar potential since in our scan at least one of the couplings of the 2HDM at $m_Z$ is different from zero. However, we can see that the larger $m_h$ masses around $160\ \text{GeV}$ require $\sin(\beta-\alpha)=1$, corresponding to the alignment well inside the decoupling limit~\cite{Gunion:2002zf,Grzadkowski:2014ada}, and hence are in the limit of the noncommutative SM Higgs scalar potential with a too heavy mass. 
In this way, we recover the rule-out prediction for $m_h$  of the SM in terms of NCG~\cite{Chamseddine:2010ud}.
Conversely, we need to stay well inside the region of 2HDM to have a consistent SM-like Higgs boson mass. In particular, we can achieve the alignment limit without decoupling~\cite{Gunion:2002zf,Grzadkowski:2014ada,Bernon:2015qea,Karmakar:2018scg,Grzadkowski:2018ohf} as illustrated below.

\begin{figure}
\centering
 \includegraphics[scale=0.7]{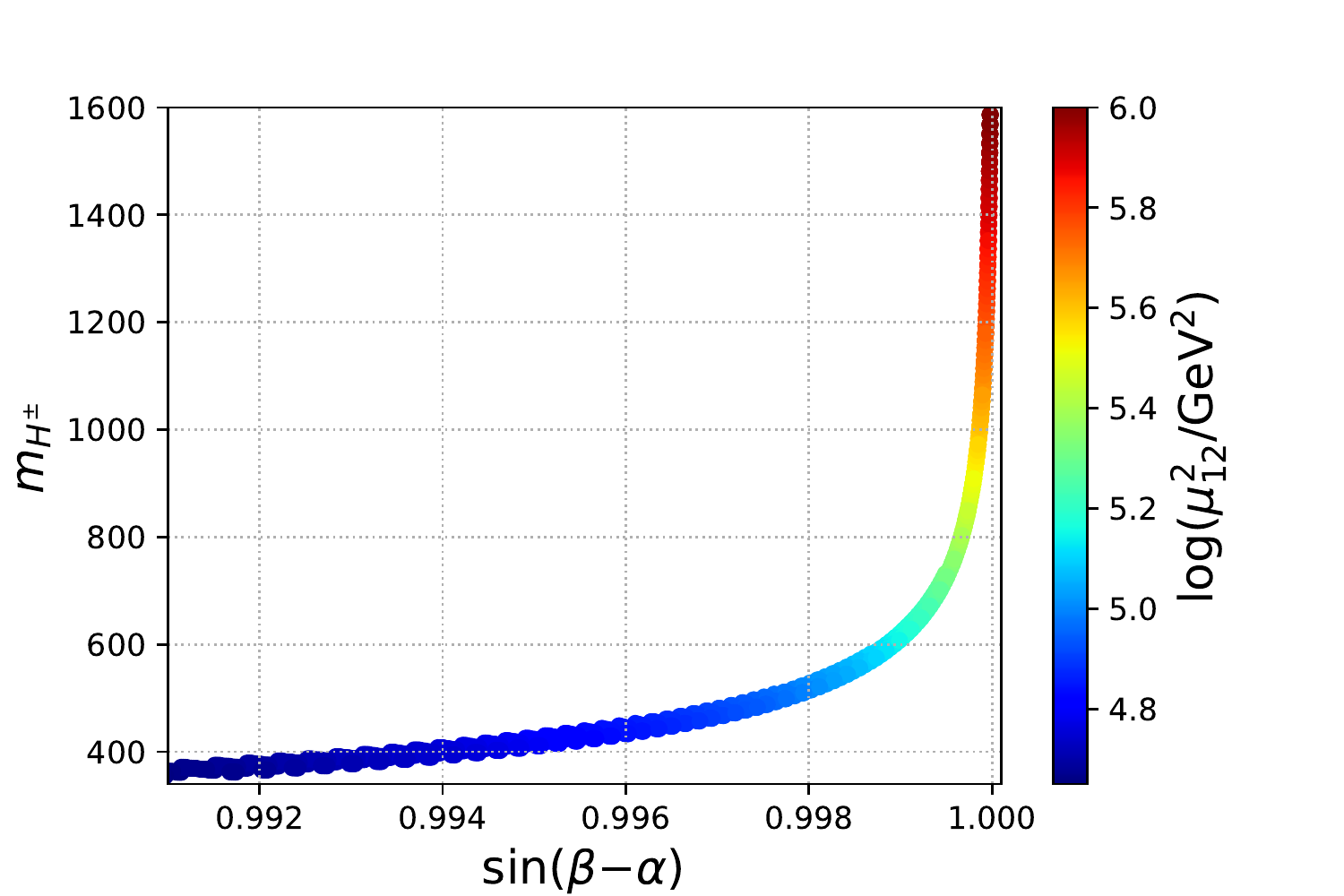}
\caption{$m_H^{\pm}$ vs $\sin(\beta-\alpha)$ with the color code indicating the value of  $\log( \mu_{12}^2/\text{GeV}^2)$. We only consider the parameter space that is fully consistent with collider searches.}\label{fig:mch}
\end{figure}

In Fig.~\ref{fig:mch}, we observe the behavior of the charged Higgs boson mass as a function of the coupling $\sin(\beta-\alpha)$, with $\mu_{12}^2$ in the color code. In this case, we only consider the parameter space that is totally consistent with the HiggsBounds constrains corresponding to the yellow region in Figs.~\ref{fig:mh} and~\ref{fig:largemh}. We see also that the non-decoupling region, $350$ GeV $\lesssim m_{H^{\pm}}\lesssim 600$ GeV, is achieved for $ \mu_{12}^2\lesssim 10^5$ GeV$^2$. 

We can see that the noncommutative 2HDM is compatible with the scalar discovered at the LHC with a value of $\sin(\beta-\alpha)\gtrsim 0.99$, which is in accordance with the alignment limit including a spectrum of non-decoupled extra scalars.

\section{Conclusions}
The noncommutative geometry approach to the standard model display a 2HDM
structure when the second-order axiom is used rather than the massless
photon condition.
By working with the type-II 2HDM as a simple realization, we are able
to recover the SM in the decoupling limit with the rule-out too heavy
SM Higgs mass.
By varying the only remaining two free parameters of the 2HDM scalar
potential with $CP$ conservation: $\lambda_5$ at the grand unification
scale $\Lambda$, and the softly broken type-II $\mathbb{Z}_2$ term,
$\mu_{12}^2$, we are able to obtain regions with a SM-like Higgs in
the alignment limit which is not excluded by collider constraints.
There, the extra scalar masses can be sufficiently low to be within reach
of future LHC runs.
Other 2HDM realizations without tree-level FCNC, could also be implemented in this framework.

\section*{Acknowledgments}
The work of DR is supported by Sostenibilidad UdeA, the UdeA/CODI Grants 2017-16286 and 2020-33177, and FAPESP funding Grant 2021/11383-4. FJ is supported by COLCIENCIAS (Doctorado  Nacional-757) and  would like to thank Shane~Farnsworth for useful discussions.
 AR is supported by COLCIENCIAS through the ESTANCIAS POSTDOCTORALES program.

\bibliographystyle{apsrev4-1long}
\bibliography{biblio}
\end{document}